# Financial climate risk: a review of recent advances and key challenges

### Victor Cardenas*

*Institute for Resources, Enviroment and Sustainability, University of British Columbia

**Abstract-** The document provides an overview of financial climate risks. It delves into how climate change impacts the global financial system, distinguishing between physical risks (such as extreme weather events) and transition risks (stemming from policy changes and economic transitions towards low carbon technologies). The paper underlines the complexity of accurately defining financial climate risk, citing the integration of climate science with financial risk analysis as a significant challenge. The paper highlights the pivotal role of microfinance institutions (MFIs) in addressing financial climate risk, especially for populations vulnerable to climate change. The document emphasizes the importance of updating risk management practices within MFIs to explicitly include climate risk assessments and suggests leveraging technology to improve these practices.

**JEL Codes-** G10, G30, Q54, Q56.

## I.  Introduction

limate change stands as one of humanity's toughest  challenges. Persisting with the current trajectory of greenhouse gas emissions Ccould escalate the severity of climate disasters, such as catastrophic floods, hurricanes, and wildfires, alongside increasing costs arising from shifts in public sector policies, innovation, and the transition to a green economy (IPCC, 2021).

The impact of climate change on the environment, ecosystems, biodiversity, and consequently, economies has the potential to threaten the global financial system and stability. Recently, it has been identified as a novel risk for financial markets. Financial climate risks, commonly known as "climate-related risks," are broadly categorized into physical and transition risks (TFCD, 2016).

This document delves into the process of defining this risk and the challenges encountered in synthesizing the impact of climate change on economies and financial markets within the context of financial climate risk. It highlights the role of microfinance institutions as a critical yet overlooked participant in the global discussion on this emergent risk.

Climate change analysis, an inherently complex phenomenon, spans several scientific disciplines. Over recent decades, scholars have rigorously examined the effects of climate variability, driven by greenhouse gas emissions, on the global climate and the economic ramifications of these changes.

The 1992 United Nations Earth Summit in Rio de Janeiro, Brazil, saw the establishment of the UNFCCC and the IPCC as its scientific advisory body, to formalise and advance a unified body of knowledge. Recently, there has been a significant increase in scholarly activity within the realms of economics and financial markets, partly due to a 2015 statement by the Bank of England (Carney, 2015) warning of climate change's potential negative impacts on these sectors. Consequently, the scientific community, regulators (FSB, 2022), financial market participants (Grippa et al., 2019), and practitioners have increasingly directed their efforts toward research specifically focused on the emerging field of financial climate risk.

Delineating financial climate risk accurately poses a significant challenge (Aven, 2020). This is largely due to the myriad questions surrounding the application of the concept of risk in climate change definitions within the scientific literature(Aven, 2016, 2010). These questions could be extended to the emerging field of financial climate risk. Notably, the concept of risk is widely utilized in research related to economics, finance, and the physical sciences that study climate change. The intersection of these fields, particularly in studies on the impacts of climate change, offers numerous research opportunities, as highlighted in this paper.

International financial market authorities have proposed comprehensive conceptualizations of financial climate risk (FSB, 2022) to capture the complexity of the challenge posed by climate change. Many theoretical issues remain at the forefront of scientific exploration, making the effective scientific address of these issues somewhat unexplored.

This paper examines an often-overlooked aspect of financial climate risk related to populations in poverty. Poverty by itself exacerbates vulnerability to climate change (Hallegatte et al., 2018) through engagement in low-yield activities, reliance on natural resources, and residence in high-risk areas. Current financial climate risk assessments do not fully consider the specific needs and constraints of this population, especially those in developing and less developed economies.

In developing and less developed economies microfinance institutions (MFIs) are more accessible than commercial banks, playing a critical role in poverty alleviation (Abrar et al., 2023; Javid & Abrar, 2015). However, financial vulnerability means that a single disaster, worsened by climate change, can obliterate assets. Therefore, MFIs should consider to update their risk management practices to address financial climate risk explicitly. This includes leveraging technology for risk management, conducting climate risk assessments, and creating financial products tailored to climate change mitigation. This area presents significant potential for growth in this sector.

This document is structured into seven sections, including the introduction and conclusions. The second section explores the relationship between the economic and financial sciences and the physical sciences of climate change. Following this, the third section addresses theoretical challenges in defining financial climate risk. The fourth section lists primary definitions of financial climate risk and their associated challenges. Subsequently, the prospects for estimating financial climate risk in its current trend are assessed in the fifth section. The sixth section discusses how financial climate risk represents a significant opportunity in the MFI sector development agenda.

## II. The link between Climate change sciences and economic and financial sciences

The complexity of climate change requires the collaboration of multiple branches of science. This section aims to explore these branches to understand and estimate future patterns and potential impacts. It examines three key areas of scientific knowledge: atmospheric sciences (Hyman, 2017; Rohli & Vega, 2018), the impact of climate (Hain et al., 2023; ISIMIP, 2023), and economics and finance (Ahmed, 2022; BIS, 2021; Campiglio et al., 2023; TDCC, 2023).

Addressing climate change necessitates a broader multidisciplinary approach beyond atmospheric sciences, incorporating physical, chemical, and biological studies to grasp the full extent of climate change impacts.

The framework outlined above highlights the intensified greenhouse effect as a direct cause of global warming. Greenhouse gases (GHGs), naturally present in the atmosphere, regulate Earth's temperature by absorbing and retaining solar radiation. However, human activities, such as burning fossil fuels, deforestation, and industrial operations, have significantly increased these gases' levels in the atmosphere, resulting in more heat retention and the warming of the Earth (Reisinger et al., 2020).

This warming is uneven and affects various parts of the climate system, altering precipitation patterns, increasing the frequency and intensity of extreme weather events, and causing sea levels to rise due to the melting of land-based ice sheets and the thermal expansion of saltwater. Atmospheric warming also impacts ecosystems, affecting biodiversity, agriculture, and water supplies (IPCC, 2021).

Thus, the complex nature of climate science is underscored by the simultaneous interactions among various elements of the Earth system. These interactions are critical in regulating the climate and intensifying warming through feedback mechanisms. Two major global modeling efforts have been initiated to enhance our understanding of climate change: 1) climate modeling, including climate change, and 2) modeling of physical and socioeconomic impacts (Amendola et al., 2013; Guin, 2017; Monier et al., 2018).

Regarding climate modeling, the World Climate Research Programme (WCRP) sponsors the development of the Coupled Model Intercomparison Project (CMIP). From its early phases (CMIP1 to CMIP3) to the latest complete study (CMIP6)(Eyring et al., 2016) and newest one (CMIP7), CMIP integrates sophisticated computer programs that generate a digital representation of Earth. The models digitize the processes and interactions among Earth's climate system components, including the atmosphere, ocean, land surface, cryosphere, and biosphere. The models are used to simulate the potential effects of future greenhouse gas emission scenarios on Earth's climate, including the extent of warming of the air and oeans and and changes in precipitaton patterns.

Regarding a second global effort, the Inter-Sectoral Impact Model Intercomparison Project (ISIMIP)(ISIMIP, 2023) aims to improve global and regional risk management by advancing knowledge of climate change risks through integrating climate impacts across sectors and scales in a multi-impact model framework. ISIMIP unifies the scenario design, and the aggregation of extreme events' effects across different sectors is part of the cross-sectorally consistent modeling framework developed under this platform.

Both analytical efforts, ISIMIP and CMIP, were catalyzed by the United Nations Framework Convention on Climate Change (UNFCCC) through the Intergovernmental Panel on Climate Change (IPCC), which serves as a scientific advisory body to the UNFCCC. The IPCC plays a pivotal role in providing comprehensive assessments of current scientific knowledge on climate change and its potential environmental and socio-economic impacts. Three different working groups has been created under IPCC, WG1 is 'The Physical Science Basis', WG2 is the 'Impacts, Adaptation and Vulnerability and WG3 is the 'Mitigation of Climate Change'.

While advancements have been achieved in the pursuit of knowledge expansion via international collaborations, obstacles remain. The estimation of climate change, for instance, world mean temperature projections or emissions by type of greenhouse gases (GHGs), are not the output of a single model. Complex interactions are needed, as described above, to estimate them.

The CMIP and ISIMIP approaches help in comprehending the historical and current aspects of climate and climate change. Nevertheless, the future takes a different methodology. Scientists worldwide agree to create future scenarios. Two types of scenarios have emerged: Representative Concentration Pathways (RCPs) and, in CMIP6, Shared Socio-economic Pathways (SSPs).

RCPs are scenarios exploring the concentration of greenhouse gases in the atmosphere. SSPs outline potential future worldwide trends in demography, economics, technology, energy consumption, and environmental regulations. Each SSP offers a narrative of potential societal evolution and they are used to evaluate the effects of climate change and the challenges in adapting to and mitigating its impacts.

SSPs and RCPs (Gibbons et al., 2021) vary primarily in their focus and application. SSPs provide detailed accounts of potential socio-economic scenarios. In contrast, RCPs concentrate on the scientific aspects of climate change, specifically examining the outcomes of greenhouse gas emissions and concentrations.

A singular all-encompassing model that elucidates the intricacies of socioeconmic feedbacks of emission pathways with climate impacts evolving with time has yet to be established in this intersection of physical and social sciences. Instead, the prevailing approach involves coordinating multiple models, institutions, and scientists to facilitate their respective development, hosting, and calibration. Generally, the results of this worldwide modeling endeavor are reflected in the evaluation of the effects of specific variables (e.g., temperature and greenhouse gas emissions) on other critical variables (e.g., precipitation, sea level, ice melting, impacted populations, or crops) using models like ISIMP or CMIP, respectively. Furthermore, to pursue the estimation of future impacts, the assessment should incorporate SSP or RCP into its analysis.

In this context, the Integrated Assessment Models (IAMs) paradigm plays a key role (Hare et al., 2018). The IAMs models, currently fed by models like CMIP and ISIMP or proxies to them, use explicit assumptions to simulate the behavior of complex systems, integrating climate and knowledge from multiple disciplines to assess the environmental, economic, and social impacts of climate change.

The macro-level IAMs approach provides a laboratory to assess the technological and economic feasibility of climate goals on a specific basis. Through simplified economic and climate science models, it evaluates different populations and economic and technological pathways, allowing an assessment of the feasibility of achieving specific climate change goals (Kandlikar & Ribsey, 1995). However, IAMs' main strength lies in assessing global trends rather than conducting accurate studies at regional or local levels.

The analysis of climate change and its impacts follows a deductive approach, moving from general to specific aspects. This involves understanding climate processes and the effects on natural phenomena and, subsequently, on the population, environment, and biodiversity.

The current challenges in the global research agenda includes to examine comprehesively the effects on the financial sector, its economy, and sub-sectors, at macro and micro level much work remains, although significant progress has been made through the analysis of Central Banks (NGFS, 2018). The economic and financial impacts methodologies are based on two main frameworks for understanding the economy at the macroeconomic level, including the macrosectorial level. The studies are based on models in the IAM style. The second framework, concerning financial impacts, represents a new field of knowledge in financial sciences, currently in development, about assimilating climate change into the body of knowledge of financial theory.

The emergence of the first framework marks a collaborative effort among central banks and members of the Network for Greening the Financial System (NGFS). This global institutional organization aims to assess the impact of climate change at the macroeconomic level and on financial systems (NGFS, 2022). Central banks are leveraging their in-house macroeconomic modeling capabilities by incorporating climate modeling techniques. The enhancement of traditional central bank modeling through this integration closely aligns with the Integrated Assessment Models' (IAMs) approach.

In the context of NGFS efforts, a set of scenarios has been developed, inspired by the Shared Socioeconomic Pathways (SSPs), but specifically tailored for the financial sector (NGFS, 2023). While SSPs and Representative Concentration Pathways (RCPs) are extensively utilized in academic and policymaking circles to understand the broad implications of socioeconomic choices and greenhouse gas emissions on climate change, NGFS scenarios aim to inform the financial industry about how different climate futures could impact financial stability and economic outcomes. This focus particularly includes policies and transitions that may emerge in response to climate risks.

However, despite the modeling capabilities inspired by the NGFS effort, the assessment's level of detail remains coarse when it comes to evaluating specific firms. The second framework, which focuses on the impact on the financial industry at the firm level, presents a significant challenge for both practitioners and academics.

Since 2017, the G7 group of industrialized countries has started recognizing climate change as a threat to financial stability (FSB, 2022). Financial regulators in each country are enhancing their analytical efforts to understand the uncertainties surrounding the impact of climate change on the performance and financing of regulated firms. The NGFS efforts serve as a foundation for regulatory analysis, aiming to improve regulations through a scientific methodology supported by central banks and financial authorities.

The nascent literature on financial markets and climate risk explores the evaluation of climate risks across various asset classes and strategies and how climate change affects company valuations. This leads to funding for climate mitigation and adaptation projects.

This body of literature has expanded significantly, especially post-2017. However, as previously indicated, efforts at the firm level lack coordination in terms of climate modeling, impact assessment, and financial stability topics.

A market for climate-related risk modeling services focused in corporations has emerged due to the implementation of a new worldwide regulatory framework focusing on climate-related risk disclosure requirements. Multiple modeling firms have conducted comprehensive studies to address this increasing need, primarily using black-box methodologies[1]. However, the confidential nature of these methodologies limits collaboration[2] among academic scientists and modelers, thereby restricting the ability to advance research globally at the firm level.

The current research (Calvet et al., 2022) shows a strong connection between climate sciences and finance, as financial markets may be influenced by various climate-related risks. The analytical efforts have been categorized into four subcategories: environmental uncertainty (examining the influence of climate variables on economic variables), economic climate risk (analyzing the impact of climate on the economy and markets), climate policy risk (assessing the risk of new regulations on business and economic performance), and financial losses (studying the effects of climate on financial asset returns, including stranded assets) that could affect the markets either directly or indirectly. Possible outcomes include disruptions in production and consumption, decreased asset value, economic damage, trade impediments, and political instability, which may jeopardize assets and financial portfolios.

In order to see the interaction of the all scientific knowledge mentioned above, see Figure 1, which illustrates the relationship between three branches of science: atmospheric sciences (referred to as "climate modeling"), impact modeling such as ISIMIP, and economic and finance sciences. The solid line in each body of knowledge represents the formality of structured and institutionalized initiatives, such as CMIP and ISIMIP, both associated with the UNFCCC and IPCC. The dotted line indicates a lack of organized effort, with just individual contributions from the academic scientific community. The G7 group of states supports the financial stability subject via the Financial Stability Board in this regard. This umbrella organization oversees modeling programs like NGFS. The most active and productive knowledge bodies are those focusing on the intersection of climate modeling and impact modeling, as well as those involved in financial stability, including regulators and regulated financial entities. Modeling for unregulated companies, especially small and medium-sized enterprises (SMEs), lacks formal study.

---

[1] For example, Baringa, Climafin "Climate Finance Alpha", MSCI "Carbon Delta", 2i Investing Initiative, ClimateWise, Ortec Finance, The Climate Service, Carbone 4, RMS Moody's, South Pole, Planetrics, Four Twenty Seven, Acclimatise, Rhodium Group, XDI Cross Dependency Initiative, Verisk, Riskthinking.ai, Climate X.
[2] Cooperation limited in regard with the level of cooperation among academic researchers in the context of IPCC and research centers and universities worldwide.

**Figure 1. Relationship between climate modeling, impact modeling and economic and finance modeling**

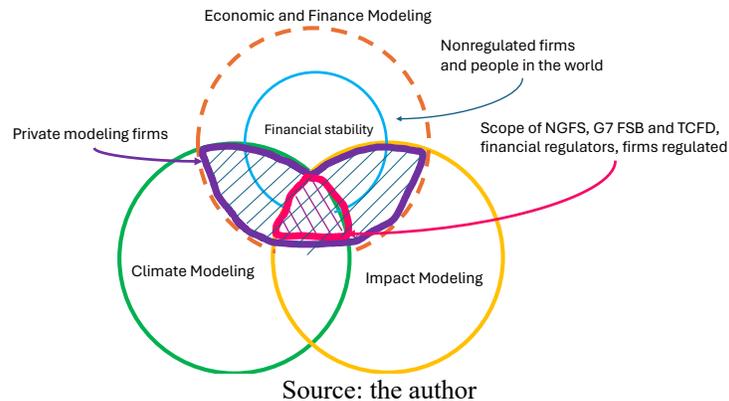

Source: the author

## III.   THEORETICAL CHALLENGES DEFINING FINANCIAL CLIMATE RISK

Defining the concept of climate change risk presents several challenges, especially when considering the intersection between financial and climate change sciences. This chapter addresses the discussions and criticisms found within the climate change literature regarding the treatment of climate change as a risk. It will also explore how these issues extend to the emerging literature on financial climate risk[3].

### III.1. RISK THEORY: AN OVERVIEW

The concept of risk is not new; references found by some authors date back more than 2,400 years (Bernstein, 1996) and others to 3200 BC (Covello & Mumpower, 1985). However, early key concepts to assess risk formally have been discussed since the 17th century (Laplace, 1820; Ore, 1960; Sheyin, 1974).

However, as Aven (2015) points out, the science of risk has been formally studied with rigorous scientific analysis from 30 or 40 years ago (Hansson & Aven, 2014; Le Coze et al., 2014; Thompson et al., 2005). For this author, the field of risk knowledge has two main aspects, the first related to the applications of theoretical concepts and the second to the analysis of the foundational concepts of risk theory. However, in the literature (Heckmann et al., 2015; Tang & Nurmaya Musa, 2011), there is a thin line between both aspects of this knowledge field. Recently, academia, through organizations such as the Society for Risk Analysis[4] (SRA), has formalized foundational concepts on risk, creating a forum for the debate on risk theory.

The SRA has issued a glossary[5] aimed at providing a consensus on the foundational risk concept. The broad definition could be summarized as "future activity with consequences on something that humans value." This concept is typically described through several metrics, expressed in terms of probabilities as a measure of uncertainty and/or expected consequences, as a measure of damage or loss, typically through an expected number or disutility, or based on specific probability distributions.

Aven (2015) discusses in detail the typical metrics used and how easily these metrics can be misused as a risk measure by themselves. He states, "For example, the expected consequences can be informative for large populations and individual risk, but not otherwise." In this regard, Aven (2015) suggests that for a metric to provide decision-making assistance for a particular choice scenario, it is necessary to identify a specific metric or collection of them that matches with the requirements and not explain the risk solely by the metric itself.

A foundational concept intrinsic to the risk domain is uncertainty. The probabilistic approach is the predominant technique for managing uncertainty. Within this context, frequentist probability is the most commonly employed tool; risk assessment, in this regard, is both epistemic and aleatory in nature. Epistemic uncertainty refers to the lack of knowledge regarding models or the consequences of uncertainty in the system or model. In contrast, aleatory uncertainty pertains to the degree of unpredictability or variability intrinsic to a given system. Conversely, Bayesian subjective probability conceptualizes probability as a degree of belief that is subjective in nature. This interpretation posits that probability serves as a means for an individual to express their level of ambiguity regarding the world. Bayesian probability updates the probability estimate for a hypothesis in light of new evidence or information using Bayes' theorem.

---

[3] In this paper, we will use the concepts of financial climate risk and related climate risk as synonyms.

[4] The Society for Risk Analysis (SRA) is a multidisciplinary, interdisciplinary, scholarly, international society that provides an open forum for all those who are interested in risk analysis.

[5] SRA (2015b). Foundations of risk analysis, discussion document. www.sra.com/resources

The link between epistemic uncertainty and Bayesian subjective probability lies in how Bayesian methods are used to model uncertainty. These methods are inherently subjective because they require the user to specify a prior belief about the state of the world, which is then updated with new evidence. This subjective prior can be seen as representing epistemic uncertainty because it reflects what is known and unknown about the system being modeled. Bayesian methods are particularly well-suited to dealing with epistemic uncertainty because they provide a formal mechanism for incorporating new information and for revising beliefs about the system. This allows for a dynamic uncertainty model that can evolve as more data becomes available or the system itself changes.

When there is a lack of reliable information, which raises doubts about the accuracy of the analysis, determining a subjective likelihood may not seem feasible. Nevertheless, assigning a subjective likelihood is always possible. The primary issue with assigning a specific probability is that it is perceived to convey more information than can be adequately justified.

The literature on the theory of risk has evolved substantially in recent decades. However, studies like (Aven, 2016) highlight the heterogeneity in approaches to theoretical work and practice when these approaches rely on principles and concepts that could mislead the decision-maker in the process of choosing among options.

In this regard, it is not enough to consider the use of probability methods for managing uncertainty as unquestionably effective merely because they are probability-based. Similarly, another critique is the reliance solely on probabilistic mean values as a risk metric without providing detailed information, such as the type of distribution and specifically the features of the tails, for instance. It is insufficient to convey to decision-makers an accurate scope of the metric. The fundamental principles of risk theory may be weakened and obscured behind complex analyses, potentially leading decision-makers astray.

### III.2. OBSERVATIONS ON THE CONCEPTUALIZATION OF RISK IN THE CLIMATE CHANGE LITERATURE

The body of literature on climate change risk is extensive. This subsection aims to focus on a particular subset of this literature, specifically those endorsed by the IPCC, which offers technical explanations and factual information regarding climate change(IPCC, 2021). The Assessment Reports (AR) have been the primary means for officially addressing the scientific aspects of climate change.

A subset of the literature in this field discusses how AR incorporates the concept of risk and the approach to uncertainty (Aven & Renn, 2015; Risbey & Kandlikar, 2007; Swart et al., 2009). The concept of risk is intrinsic to the definition of climate change. This subset of literature has examined the IPCC's approach to risk and uncertainties, with some supporting it and others questioning it.

The IPCC's approach is the result of a collaborative effort among several authors who, despite coordination, faced continuous challenges in aligning ideas and interpretations. Comments on this topic of risk and uncertainty have emerged, spanning from AR1 to AR4.

Through Manning (2014), the IPCC released a report explaining and validating the approach regarding AR3, addressing several topics. Among them, two key issues were "likelihood" and "confidence" as alternative ways of expressing uncertainty.

Jonassen & Pielke (2011) examine the treatment in AR4. It highlights inconsistencies in the conveyance of uncertainties across the different Working Groups (WGs), emphasizing the need for improved consistency, clarity, and systematic treatment of uncertainties in future IPCC assessments.

Budescu et al., (2009) examine IPCC reports and propose improvements. The study suggests that the method used by the IPCC may convey too-high imprecision levels and that many probabilities may be interpreted less than intended by the authors. Additionally, there is a need to focus more on the structural uncertainties that permeate the models and their projections, as well as to highlight the sources and magnitude of these uncertainties. The ultimate goal is to enhance the understanding of the findings and assessments, thereby improving the foundation for sound policy decisions.

Major challenges are found in the concepts of "likelihood" and "confidence." For instance, Risbey & Kandlikar, (2007) examine the evolution of uncertainty communication in the IPCC reports, focusing on the expressions of likelihood and confidence. It discusses the transition from subjective, linguistic categorizations of uncertainty to more formal and probabilistic classifications. The document highlights the potential for confusion when combining likelihood and confidence metrics, especially when low confidence levels are paired with very high or low likelihoods. It proposes a flexible approach to likelihood expression, matching the precision to the level of understanding. The paper also discusses the use of sequential processes for communicating uncertainty at a level appropriate to existing scientific understanding.

In response to critiques on uncertainty treatment found in IPCC reports, Morgan et al. (2009) synthesize and communicate the current state of understanding about the characteristics and implications of uncertainty related to climate change and variability. The document provides guidance on best practices for describing and analyzing uncertainty in climate-related problems, with the goal of informing public debate, policy, and operational decisions.

Finally, as consequence of critiques from several sources the IPCC issued a guideline for authors in 2020 (Reisinger et al., 2020), it aims to promote a more consistent and transparent application of the concept of risk across Working Groups in the sixth assessment cycle and to provide the foundation for its use and possible further evolution in future assessment cycles. The guidance note seeks to maximize consistency among Working Groups while recognizing differences in how the concept of risk is used across scientific disciplines and relevant literature. It clarifies the use of the term "risk" and provides an annotated definition, emphasizing the need for careful use of IPCC-calibrated uncertainty and confidence language to transparently and clearly describe adverse consequences and their potential. The document also addresses the use of risk in different contexts, such as flood risk, food security, and risk in the investment and finance literature. Additionally, it highlights the importance of considering socio-economic and behavioral factors and trends in assessing risks related to climate change. The concept of risk is presented as a key aspect of how the IPCC assesses and communicates potential adverse impacts of, and response options to, climate change, with the aim of informing decision-making and risk management.

III.3. Observations on the conceptualization of risk in the early literature on economics impacts by climate change

One of the earliest analytical efforts to examine the impact of climate change risk on the economy was the Stern Review (Stern, 2006). The Review describes climate change as the most significant and widespread market failure ever observed, presenting a unique and challenging problem for economics.

The Review sparked considerable discussion and critique, particularly regarding the methodologies used for estimation and conclusions. A key area of criticism was the treatment of uncertainty. In this context, the uncertainty about future consumption could be addressed by adjusting the discount rate. However, critics like William Nordhaus argued that the Review's application of a low discount rate was inappropriate. A low discount rate increases the present-day significance of future costs and benefits. Conversely, it was noted that economic agents' approach to risk aversion and the emphasis on catastrophic scenarios tend to undervalue events with low likelihood.

This Review led to widespread debate and commentary on the methodologies employed to make estimates and draw conclusions. The handling of uncertainty was a particular point of contention. Regarding this issue, the ambiguity surrounding future consumption might be managed by altering the discount rate. Nevertheless, critics such as William Nordhaus have argued that the Review's adoption of a low discount rate was unsuitable. A lower discount rate elevates the relevance of future expenses and benefits in current terms. On the flip side, the tendency of economic actors to be risk-averse and to emphasize catastrophic scenarios results in giving less weight to events with low probabilities.

The Stern Review initiated a vast body of literature (Cole, 2008; Peacock et al., 2014) discussing the concepts, assumptions, and modeling approaches regarding climate change from an economic perspective. Its primary contribution to the global discourse on climate change was framing it as an economic issue, not merely an environmental concern, by quantifying the potential costs of inaction in clear economic terms. However, as a communication piece on the impact of climate change on the global economy, the document reproduces some inaccuracies related to risk and uncertainty concepts originally developed in IPCC reports.

In a search for the concept of risk in the Stern Review, it was generally found to be consistent with the usage of the risk concept in IPCC reports, albeit with some inadequacies. For example:

1) "The scientific evidence points to increasing risks of serious, irreversible impacts from climate change associated with business-as-usual (BAU) paths for emissions." (Stern, 2006, p. iii)

2) "Under a BAU scenario, the stock of greenhouse gases could more than treble by the end of the century, giving at least a 50% risk of exceeding a 5°C global average temperature change during the following decades." (Stern, 2006, p. iv)

3) "Estimating the economic costs of climate change is challenging, but there are a range of methods or approaches that enable us to assess the likely magnitude of the risks and compare them with the costs.… Warming will have many severe impacts, often mediated through water: Melting glaciers will initially increase flood risk and then strongly reduce water supplies, eventually threatening one-sixth of the world's population, predominantly in the Indian subcontinent, parts of China and the Andes in South America." (Stern, 2006, p. vi)

The concept of risk is used heterogeneously throughout the Report, as seen in the examples above. In (1), risk is associated with likelihood or chance, adding a sense of magnitude to the likelihood. In (2), risk highlights a probability linked to a threshold. In the context of (3), risk encompasses more than just the likelihood of an event occurring; it also includes the various effects associated with flooding. This scenario involves a complex system that includes the interplay between the costs, melting glaciers, and the risk of floods.

The Stern Review, similar to the IPCC reports, views risk as the outcome of the likelihood of an event occurring and its resulting impact, resembling an expected value, as highlighted by (Aven & Zio, 2014). The literature has debated whether this is an incorrect use of the risk concept since it assumes an expected value without explicitly stating it. However, relying solely on probabilities without considering other factors may lead to potential errors. This is because probabilities, as standalone figures, do not convey information about potential variances. Within the report, the notion of probability is not explicitly defined but is presented as an ambiguous and imprecise concept.

### III.4. OBSERVATIONS ON THE CONCEPTUALIZATION OF RISK IN THE FINANCIAL CLIMATE RISK LITERATURE

The literature on financial climate risk is still at a very early stage at a theoretical level; in part because there is no body of knowledge that unifies the multidisciplinary components involved, as does the reports linked to the UNFCCC through the IPCC publications.

The economic and financial sciences have largely contributed to the development of knowledge of the concept of risk both theoretically and empirically. However, when this knowledge is combined with the knowledge of climate change generated by the physical sciences, compiled by the IPCC, there is still no theory that connects all bodies of knowledge under the same umbrella.

However, the efforts that exist are given after the release of the TFCD recommendations in 2017 (TFCD, 2016). Such efforts are in two aspects, the first is related to the financial disclosure documents to authorities and investors in which the concept of risk is mentioned, which in general are corporate reports of companies, where they voluntarily disclose information concerning the climate risks faced by the companies issuing the reports, following the TFCD guidelines. The corporate reports issued had an outstanding growth, according to FSB (FSB, 2022), "the percentage of companies disclosing TCFD-aligned information continues to grow, but more progress is needed. For fiscal year 2022 reporting, 58% of companies disclosed in line with at least five of the eleven recommended disclosures – up from 18% in 2020."

Thus, the most widespread use of the risk concept in the financial climate risk literature is in corporate reporting. For example, to illustrate its use, consider the voluntary reports issued by City Development Limited based in Singapore, Vodafone and Sunway, an industrial conglomerate based in Malaysia.

| Report Name | Reference in the report | Comment related to the use of the concept of risk |
|---|---|---|
| City Development Limited Integrated Sustainability Report (2021)[6] | Page 23-27 | Risk is associated with likelihood or chance as well as risk highlights a probability linked to a threshold |
| Vodafone TCFD Report (2021)[7] | Pages 10 | Risk encompasses more than just the likelihood of an event occurring; it also includes the various effects associated to specific events |
| Sunway Sustainability Report 2020[8] | Page 13 and 26 | Risk highlights a probability linked to a threshold, which include side effects linked to the event. |

The analysis of the reports allows to determinate the degree of adherence to TFCD's recommendations, however, in the three cases and probably in the majority of the reports issued, they do not mention the assumptions, models, analysis on which is based the risk estimation. The purpose of the communication piece is to use a language accessible to the greatest number of investors, at the same time, it is unacurate and limited in the analytical description of what they mean by risk in a rigorous manner.

On the other hand, the second aspect is in academic research, for example, according to the Journal of Corporate Finance, the Journal of Financial Economics, Annual Review of Financial Economics and Nature Climate Change, in various special issues of finance linked to climate change. The academic research published, the topics in which the literature has expanded are: efficiency of the market pricing of climate risks, causal links between global warming and firm or industry cash flows, venture capital, crowdfunding, and alternative finance for clean-tech, green bonds and green financial institutions, valuation of stranded assets in energy and carbon intensive industries,

---

private equity and project finance for renewable energy investments, agency conflict, board oversight, and corporate governance incentives in dealing with climate risks, institutional investors and climate change, green lending and green stress testing, hedging climate risks, corporate disclosure regarding carbon and climate risk exposures.

The following table provide a syntesis of the use of the risk concept in the each research paper.

| Author | Reference in the paper | Comment related to the use of the concept of risk |
|---|---|---|
| Stroebel & Wurgler, (2021) | 487 - 488 | Identify regulatory risk as the top climate risk to businesses and investors over the next five years, but they view physical risk as the top risk over the next 30 years. |
| Rao et al., (2022) | Page 2 | Provides a specific perspective of risk, the study draws on the 'salience theory of choice under risk', which predicts that firms pursue differential investment strategies based on differential saliency encountered by the managers. |
| Huang et al., (2021) | Page 4 | Develop a model characterized by a climate policy risk that arises from the enforcement of specific environmental standards. The concept of risk is associated to the lower firms' profitability, induced by additional costs, and to the higher excess premium to cover the costs of default by the entrepreneurs facing such lower profitability, thus triggering to default. |
| Javadi & Masum, (2021) | Page 1 | risk is associated with likelihood or chance as well as risk highlights a probability linked to a threshold, in this case how climate risk affects firms' cost of capital |
| Hickey et al., (2021) | Page 1 | risk is associated with likelihood or chance as well as risk highlights a probability linked to a threshold, in this case early obsolescence (impairment) of fossil fuel fuel power plants |
| Calvet et al., (2022) | Page 1-2 | Risk is associated with likelihood of losses in the financial sector because the climate change |
| Giglio et al., (2021) | Page 2 | The approach to the concept of risk is the impact in the prices of financial assets due to climate change. |
| Campiglio et al., (2018) | Page 465 | risk is associated with likelihood of impact to an economy, specifically the financial sector. |
| Dietz et al., (2016) | Page 5 | Risk is linked the impact in the prices of financial assets because climate change, specifically through a measure proposed as "climate value at risk". |
| Krueger et al., (2019) | Pag | The concept of risk is linked to changes in the prices of financial assets because climate change, from the point of view of investors. |

In summary, contrary to the communication pieces related to climate change oriented to mass readership, in the case of scientific research, in general terms is more precise in the use of the concept, using detailed descriptions of the meaning of risk.

## IV. DEFINING FINANCIAL CLIMATE RISK

The study of risk as a science is relatively new, with most of its theoretical development driven by advancements in practical analysis across a wide range of scientific disciplines (Aven, 2016). Climate change poses risks not just to the environment, ecosystems, biodiversity, and economies but also to financial markets (Hjort, 2016). These markets play a crucial role in the growth of economies, so any changes within them may have a ripple effect that could potentially harm the well-being of current and future generations. The field of climate change risk is now emerging as a new opportunity to expand our knowledge of risk within financial sciences.

The first time a financial authority highlighted the concern of a real threat of climate change to the financial system was in a September 2015 speech by the Governor of the UK Central Bank, titled "Breaking the tragedy of the horizon - climate change and financial stability" (Carney, 2015). The speech emphasized that climate risk is not only a scientific concept but also a part of the scope of financial risks. Consequently, the primary outcome of this speech was the concept of financial climate risk, which was, for the first time, incorporated into the field of financial sciences.

A cursory review of recent literature reveals significant impacts on financial assets, for example, in the credit market. Climate change presents considerable risks to this market globally. Studies such as those by (Nie et al., 2023) suggest that severe climate and environmental disasters can lead to an increase in non-performing loans (NPLs) within the banking sector. Other financial entities like insurance companies, asset managers, and stock and debt issuers are also vulnerable to climate change risks. For instance, climate events can cause physical damages, economic losses, and disruptions that impact the solvency and profitability of financial institutions. Conversely, as a result of climate change, the default risk, which is the possibility of financial entities going bankrupt, can increase (Iqbal & Nosheen, 2023). This could jeopardize global financial stability (FSB, 2022).

This new field of financial science grapples with the challenge of accurately naming the risk climate change poses to financial markets. The term widely adopted in the literature is "climate-related risk" (TCFD, 2017). However, this designation does not explicitly link to financial markets. Initially, in the TCFD's report to the FSB (2016), the term used was "climate-related financial risk," a phrase that has since fallen out of favor, being supplanted by "climate-related risks." For clarity in this document, we will use the term "financial climate risk" to refer to climate-related risk.

Another challenge posed by this emerging risk is in accurate definition. Given the novelty of this risk, questions have been raised about whether it should be categorized within the scope of financial sciences. Some literature addresses this relationship explicitly as a financial risk, while others implicitly assume it to be so. However, this debate masks a deeper issue regarding whether financial climate risk ought to be considered a part of the science itself.

Below, it is discussed several definitions on financial climate risk from international authorities that provide regulatory guidance to the global financial system.

Table 1

| Source | Reflexion on the definition |
| --- | --- |
| TCFD (2020) | Climate-related risks refer to the potential adverse impacts on businesses and financial systems resulting from climate change and the transition to a lower-carbon economy. These risks encompass physical risks, such as extreme weather events and changes in climate patterns, and transition risks, including policy changes, reputational impacts, and shifts in market preferences and technology. |
| FSB (2022) | Compile several definitions through the following common elements in existing definitions: component 1) A physical risk definition that includes both acute and chronic risks. Component 2) A definition of transition risk includes technological developments, behavioral and social change, and policy changes. Component 3) A definition of liability risk: Liability risk associated with physical and transition risks, such as potential financial losses stemming directly or indirectly from legal claims. |
| NGFS, (2018, 2022) | According to the Network for Greening the Financial System (NGFS), climate-related risks are defined as risks that arise from climate and weather-related events, such as droughts, floods, storms, and sea-level rise, as well as from progressive shifts in climate and weather patterns, such as increasing temperatures. These risks can directly and indirectly impact financial institutions, including reduced value of assets, increased insured damages, disruption of business operations, and potential macroeconomic impacts. Additionally, transition risk is also considered, which is the financial risk resulting from the adjustment process towards a lower-carbon economy prompted by changes in climate policy, technology, or market sentiment. The NGFS emphasizes the importance of understanding and managing these risks to ensure the resilience of the financial system. |
| BIS, (2021) | In a literature review, IBS found that Climate-related risks refer to the potential adverse impacts and uncertainties arising from the changing climate, including physical and transition risks. Physical risks encompass acute hazards such as rising sea levels, floods, storms, and heat waves, which can lead to property damage, supply chain disruptions, and increased insurance claims. Transition risks are associated with the shift towards a low-carbon economy, including policy changes, technological advancements, and market shifts, which can affect the value of assets and the financial stability of institutions. These risks can have implications for various sectors, including banking, insurance, real estate, and government finances. They may require proactive management and adaptation strategies to mitigate their impact. |
| ISSB, (2023) | Risks stemming from climate change can be categorized as event-driven (acute physical risk) or stemming from longer-term shifts in climate patterns (chronic physical risk). Acute physical risks emerge from weather-related events like storms, floods, droughts, or heat waves, which are growing in severity and frequency. Chronic physical risks result from gradual shifts in climate patterns, leading to sea level rise, diminished water availability, loss of biodiversity, and alterations in soil productivity.

Such risks may have financial repercussions for entities, including costs from direct asset damage or indirect impacts like supply chain disruptions. An entity's financial performance may also suffer due to changes in water availability, sourcing, and quality, as well as extreme temperature variations impacting premises, operations, supply chains, transportation requirements, and the health and safety of employees. |
| IMF (Beschloss & Mashayekhi, 2019) | Climate-related risks refer to the potential negative impacts on the economy and financial system resulting from the physical and transition effects of climate change. Physical risks include damage to property, infrastructure, and land, as well as extreme weather events and gradual climate changes, leading to business disruption, asset destruction, migration, and the need for reconstruction or replacement. Transition risks arise from changes in climate policy, technology, and consumer preferences during the adjustment to a lower-carbon economy, affecting the value of financial assets, business models, and the stability of the financial system. These risks can lead to |

| | increased default risk of loan portfolios, lower asset values, and higher insurance claims, impacting the stability and functioning of financial institutions and markets. |
|---|---|
| IAIS, (2021) | Risk posed by an insurer's exposure to physical, transition, and/or liability risks stemming from or related to climate change. |

Currently, the discussion surrounding financial climate risk is focused on two key aspects: 1) theoretical implications and approaches to risk estimation (academia and practitioners), and 2) From a pragmatic standpoint (regulators, financial authorities and private modeling firms), it emphasizes the disclosure of information through regulatory processes. It could be argued that comprehending and modeling the risks prior to requesting information that can help in understanding financial climate risk is an intuitive course of action. However, in the real world, where both aspects co-exist and sometimes without noticeable interaction.

IV.1. DISCUSSING THE THEORETICAL IMPLICATION OF FINANCIAL CLIMATE RISK DEFINITION

Generally, the definitions provided align closely with the original recommendations of the TCFD (TFCD, 2016), which break down into two main concepts: physical risk and transition risk.

In summary, physical risks refer to the direct consequences of climate change, including severe weather events, rising sea levels, and shifts in weather patterns. These hazards may be classified as acute, resulting from discrete weather events, and chronic, stemming from longer-term changes in climate patterns (TFCD, 2016).

Conversely, transition risks pertain to the challenges of adapting to a lower-carbon economy. This includes potential hazards associated with changes in policies, advancements in technology, market conditions fluctuations, and societal responses and adjustments to climate change. These risks emerge from global efforts to mitigate greenhouse gas emissions and/or adapt to the impact of climate change. They can have significant financial and operational consequences for companies and economies as they move toward more environmentally friendly practices.

Originally, the TCFD's recommendations were designed to "encourage organizations to evaluate and disclose, as part of their annual financial filing preparation and reporting processes, the climate-related risks and opportunities most pertinent to their business activities. (TFCD, 2016, 2023)" The distinction between these two types of risks aids in clarifying the process for public disclosure, as it presents both perspectives on the impact of climate change on organizations and economies, respectively.

In fact, this approach has subsequently been followed in the literature, which differentiates between the two risks and analyzes them separately. The divisions are implicit within each risk, categorized into acute and chronic physical risks and transition risks divided into legal-policy-and-litigation, technological, market-economic, and reputation risks.

However, upon careful analysis, each of these risks can be distinguished by a subtle boundary, which can often be difficult to discern since the divisions are ultimately arbitrary. In this context, a new subset of literature has recently emerged, assessing climate risk in a more comprehensive way, as experienced in the real world. This recent work describes complex climate change risks, considering concepts of compounding, connecting, and cascading interactions among them (Brovkin et al., 2021; Lawrence et al., 2020; Naqvi & Monasterolo, 2021; Schlosser et al., 2023; Urbani-Ulivi, 2019).

For example, (Simpson et al., 2021) present a framework consisting of three categories of increasing complexity. Category 1 focuses on interactions among a single driver for each determinant of risk, such as hazard, exposure, vulnerability, and response. This level of assessment considers the interactions within a single risk domain, providing a detailed understanding of the specific interactions that generate risk. Category 2 involves interactions among multiple drivers within each determinant of risk. This level of assessment expands the scope to consider how multiple factors within each risk domain interact, providing a more comprehensive understanding of the complexity of risk. Category 3 encompasses interactions among multiple risks. This level of assessment considers the interconnected socio-economic, environmental, and technological systems that generate climate change risk. It also takes into account the interactions and trade-offs between different risks, providing a holistic view of the complex climate change risk landscape.

Figure 2 illustrates the multiple material and conceptual boundaries across which interactions can dampen or amplify climate change risks. It provides examples of these interactions, including cross-sectoral interactions such as those between water, energy, food, and health; temporal lags such as those between climate extremes and behavior change; spatial telecoupling such as for food trade networks and breadbasket failures; and interactions of multiple mitigation and adaptation response options such as urban greening and fossil-fueled air conditioning as responses to extreme heat. The figure aims to demonstrate the complexity and interconnectedness of climate change risks, highlighting the need for a comprehensive approach to risk assessment that considers the interactions across various domains.

**Figure 2. Conceptual boundaries across which interactions can dampen or amplify climate change risks**

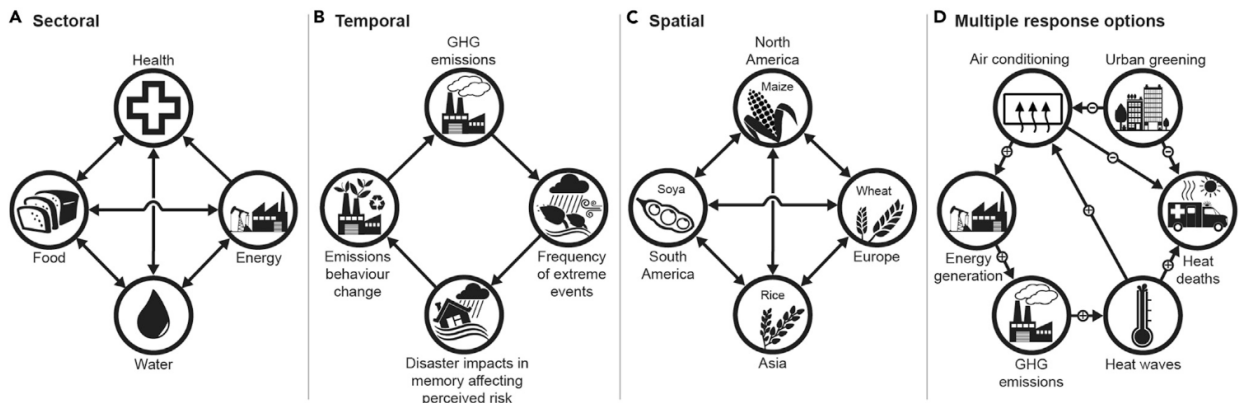

Source: taken from Simpson et al. (2021)

In recent years, the literature (Clarke et al., 2018) has shifted from conceptual models, which needed to orient research towards models considering risk composition, to methodologies and models that actually do so. Formally, the need to move in this direction was emphasized by IPCC, (2012), Leonard et al., (2014) and recently (Zscheischler, 2020).

Recent studies have shifted their analytical focus from examining a single hazard to exploring two or three hazards simultaneously within the same model. For instance, (Zscheischler et al., 2020)) conducted an analysis of dry summers, incorporating both temperature and precipitation for the first time. (Aghakouchak et al., 2020) was the first to estimate the likelihood of concurrent drought and heat waves, drawing on data from the California drought of 2014. (Van Den Hurk et al., 2015) examined compound surge and precipitation events. (Martius et al., 2016) analyzed compound events involving precipitation and wind extremes. Stalhandske et al., (2024) focused on compound tropical cyclone and flooding events.

Research in compounding risk has evolved to encompass multiple systems and economic sectors. Schlosser et al., (2023) introduce a novel approach to compounding risk, aiming to develop a risk-triage platform. This platform, known as the System for the Triage of Risks from Environmental and Socio-Economic Stressors (STRESS), facilitates the assessment of compounding risks across various systems and sectors. It is designed to offer a rapid, low-cost, screening-level assessment to identify "hotspots" of co-existing, co-evolving situations and to inform where resources and efforts can be most effectively allocated. The platform aims to characterize the extent to which various risk factors co-exist, compile metrics that quantify these risk factors in a flexible and combinable manner, identify how the aggregate risk landscape changes when individual risk factors are combined, and enable users to explore various risk-factor combinations at their discretion. Intended to prioritize efforts in vulnerable regions, the platform seeks to enhance resilience against potential compounding stresses and provide a quantitative basis for identifying risk "hotspots." Additionally, the platform is designed to be flexible and extensible, allowing for the augmentation or modification of metrics in response to feedback or requests from the research community, as well as government and private stakeholders.

Within this framework, the NGFS has emerged as a comprehensive endorsement by international governing bodies responsible for overseeing the global financial system. It leverages the expertise of technical teams from central banks within the organization, who have declared their intention to focus on addressing complex risks. While preserving the structure of the initial TCFD recommendations, it categorizes risks into physical and transitional.

The level of implementation concerning the inclusion of compound shocks in scenario analysis varies significantly across countries. In July 2023, a survey conducted among NGFS members found that only 25% of respondents had incorporated compound climate shocks into their climate-related scenario analyses. However, about 60% reported that considering compound shocks was a regular practice in their general scenario analyses and stress testing, which did not specifically focus on climate-related scenarios. The survey also revealed that a significant number of central banks and regulators are now integrating compound shocks into their climate scenario research and stress testing. Nevertheless, there are challenges to fully integrating compound shocks, including the lack of empirical data, methodologies, and models capable of capturing the non-linear impacts of compound climate shocks. The survey further highlighted the need for model advancements to effectively capture compound shocks and the importance of strong collaboration between academics and practitioners to integrate methodological improvements into the toolkits of central banks and supervisors. Overall, there is broad consensus on the necessity of considering compound risks when analyzing the effects of climate change. Moreover, there is a need for a realistic set of tools to help integrate compound risks into climate scenario analysis, which remains on the long-term agenda even for the more advanced technical teams of member countries.

## V.    Methodological estimation of financial climate risk

The estimation of transition and physical risks does not start from scratch. Physical risk is based on the catastrophe (cat) modeling of the reinsurance industry (Grossi & Kunreuther, 2005). Transition risk, on the other hand, is based in economics through the agent-based macroeconomic models, e.g., (Hałaj, 2018) and general equilibrium models for modeling transmission channels of monetary and fiscal policies.

### V.1.   Estimating physical risk

According to Grossi & Kunreuther (2006), a catastrophe model is a tool used to assess the risk of natural disasters and make informed risk management decisions. It consists of four basic components: hazard, inventory, vulnerability, and loss. The model first characterizes the risk of natural hazard phenomena, such as earthquakes or hurricanes, by quantifying their potential impact. Then, it identifies and characterizes the properties at risk, including their location and other relevant factors. The vulnerability of these properties to natural hazards is quantified, considering potential damage and losses. Finally, the model evaluates the potential direct and indirect losses, such as repair costs and business interruption impacts (Baxter et al., 2020; Cossette et al., 2003).

The output from the catastrophe model provides valuable information for insurers and reinsurers to manage their risk. It aids in estimating premiums for insurance coverage, tailoring coverage to reduce the risk of insolvency, and developing new strategies for managing portfolios to avoid unacceptable reductions in surplus (Guin, 2017). The model's output also assists in making decisions about risk management strategies, such as mitigation, insurance, reinsurance, and catastrophe bonds.

Overall, catastrophe models play a crucial role in assessing and managing the risk of natural disasters, providing essential information for stakeholders to make informed decisions about protecting against catastrophic events.

The main difference between catastrophe modeling and physical risk modeling lies in their fundamental approaches and objectives (Turner, 2023). Catastrophe models are primarily focused on pricing and managing natural catastrophe insurance risk in the present, relying on historical data and statistical techniques to estimate potential impacts. These models are constructed from and calibrated to historical observations, (therefore potentially omitting climate change), and are often based on statistical methods rather than dynamical ones.

On the other hand, physical risk modeling, particularly in the context of climate change, aims to understand and forecast how extreme weather might evolve in the future. It involves climate models that incorporate the mathematics of the atmosphere, interactions between components, and the impacts of greenhouse gases. These models attempt to provide a view on the distribution of potential outcomes and their likelihood, considering the impact of climate change over near-term and long-term time horizons. Physical risk modeling also seeks to account for changes in the built environment and the potential multiplier of internal variability.

### V.2.   Estimating transition risk

Agent-based modeling (ABM) is a well-known method in the social sciences (Salgado & Gilbert, 2013). This modeling approach is based on autonomous and heterogeneous agents interacting within an environment to identify the mechanisms that give rise to macroscopic phenomena of interest. For example, in macroeconomics and monetary policy, an ABM could be designed to capture the complex interactions between heterogeneous agents in the financial market and analyze the channels through which funding shocks can propagate across the system. This kind of modeling provides a comprehensive framework for analyzing the systemic implications of funding shocks in the financial system, considering the interactions between liquidity and solvency conditions and the complex dynamics of the market.

On the other hand, General Equilibrium Models (GEMs) (Carattini et al., 2021) are rooted in neoclassical economics and analyze supply and demand in different markets to determine an equilibrium point where supply and demand are equal across the entire economy. GEMs operate under the assumption that agents, including consumers and firms, are rational and have complete and accurate information. They are often used for policy research and forecasting in the field of economics, particularly for examining the overall effects of changes, such as alterations in tax laws or trade agreements.

By combining the macroeconomic insights of General Equilibrium Models (GEMs) with the micro-level complexity and flexibility of Agent-Based Models (ABMs), researchers can develop more complete and nuanced models. An integration of both macro-level equilibrium results and micro-level behavioral dynamics may provide a more comprehensive understanding of how individual choices and interactions contribute to macroeconomic patterns.

For some central banks, the modeling standard for assessing transition risk is through the Emissions, Prediction, and Policy Analysis (EPPA) model by MIT (Paltsev et al., 2005). The MIT-EPPA is an advanced economic model that analyzes the global economy. It is based on the Global Trade Analysis Project (GTAP) dataset and includes statistics on greenhouse gas and urban gas emissions. The model monitors emissions in relation to economic activity and involves enterprises making choices that minimize costs over time. The MIT-EPPA model depicts the global economy by considering many nations, regions, and related industries. Furthermore, the model encompasses a comprehensive range of technologies, including both conventional fossil fuels and more sophisticated alternatives such as bioenergy with carbon capture and storage.

## VI.   Missing voices in the process of defining financial climate risk

One overlooked aspect in the discussion of financial climate risk is the situation of populations living in poverty, who are notably more vulnerable to climate change due to several factors. First, households in poverty often engage in low-risk, low-return activities because of uninsured weather risks, which perpetuates their poverty. The threat of weather shocks can significantly reduce incomes and exacerbate the impacts of disasters. Additionally, individuals living in poverty are more likely to reside in marginal areas and rely on natural resources, heightening their susceptibility to environmental impacts. Studies (Sawada & Takasaki, 2017) have shown that these populations are more exposed to high temperatures and are likelier to work in outdoor occupations, increasing their vulnerability to extreme heat. Moreover, due to lower costs, individuals in poverty are more inclined to settle in areas prone to climate-related hazards, increasing their exposure. The limited access to financial instruments and social protection further contributes to their vulnerability.

In this context, a critical aspect often overlooked in assessing financial climate risk involves those living in poverty, their access to financing instruments, and the surrounding financial institutions (Hermes & Lensink, 2007). Particularly, there's a notable absence of emphasis on developing risk assessment methods tailored to their unique needs and constraints, especially methods that include developing and less developed economies.

### VI.1. Poverty and risk financing

One of the primary risk factors threatening those in poverty is the occurrence of extreme events resulting from weather phenomena which may be exacerbated by climate cahgne. The existing research on this issue has mostly recommended understanding and proposing strategies to reduce the effect of severe occurrences on the human and physical capital stock.

In this regard, in the literature , there is a growing interest in studying (Olivieri et al., 2011) how the population in poverty in developing economies is more vulnerable than the population in poverty in high-income economies. For example, (Hallegatte & Rozenberg, 2017) conducted a survey of 1.4 million households, representing 1.2 billion households and 4.4 billion people in 89 developing economies. The Survey analysis concludes that 26 million fewer people would be in extreme poverty if all disasters could be prevented next year.

Regardless of the impact on the value of assets damaged, natural disasters can significantly affect the well-being of asset owners. Other dimensions of disasters not captured in the estimation of losses through replacement value include impacts on health, education, income, employment, food security, nutrition, and quality of life (Hallegatte & Rozenberg, 2017; Hallegatte & Walsh, 2021).

Additionally, due to their own circumstances, the poor face several challenges in managing risk, such as imperfect or nonexistent factor markets (including labor, land, credit, insurance, and liquidity)(Sawada & Takasaki, 2017). These challenges can be magnified by natural extreme events, potentially deepening poverty traps and perpetuating the condition.

On the other hand, populations in poverty within developing economies are actively dealing with disaster risk through informal mechanisms in the shadow economy, for example, (Mendelsohn et al., 2006) assess country-specific market impacts of climate change; (Tol, 2002) estimates of the damage costs of climate change; (Hope, 2006) study the marginal impact of CO2; (Mendelsohn et al., 2006) study the distributional impact of climate change on rich and poor countries; (Olivieri et al., 2011) estimate the economic effects of climate change, and (Nordhaus, 2014) estimates the social cost of carbon: concepts and results from the DICE-2013R model and alternative approaches.

Low-income farmers have historically managed risk in agricultural production by diversifying crops, intercropping, making flexible production investments, using well-established, low-risk technologies, and entering into specific contracts, such as sharecropping. In urban areas, forming long-term business relationships based on ethnicity or affinity is widespread in the commercial and industrial sectors to mitigate risks. Generally, low-income individuals tend to have precautionary savings in the form of livestock, grains, and cash to cover the costs of extreme events ((Fafchamps et al., 1998); (Lee & Sawada, 2010)). On the other hand, various types of credit, such as lending linked to personal or commercial affiliations, money lenders, and loan sharks, are less used, while savings among family and friends are commonly utilized (Mansell, 1995).

In some developing economies, risk management techniques for the poor have evolved into formal financial instruments, such as lending mechanisms (microcredit), insurance (microinsurance), and mutual insurance. The literature in this area has significantly expanded over the last decades, assessing institutions like Grameen Bank's model, highlighting two different kinds of services, namely, credit, savings, and insurance, which could take various contractual forms, depending on how low-income households are reached.

Microfinance Institutions (MFIs) are considered a key tool for mitigating poverty and promoting development ((Javid & Abrar, 2015); (Bangoura, 2012); (Khavul, 2010); (Hermes & Lensink, 2007); (Mahmood et al., 2016)). In many economies, MFIs are the closest mechanism for accessing financing through formal institutions. (Abrar et al., 2023) found significant differences compared with commercial banks, showing that MFIs stand out in terms of higher financial intermediation, non-interest income (fee income), wholesale funding, and liquidity.

However, for major disasters, the lack of resilience and diversification inherent in the financial fragility of the poor could, in a single event, wipe out all types of assets or value reserves. This is especially true when access to MFIs is limited, and these institutions are less diversified regionally. In this context, the impact of climate change on disaster risk is the last straw, magnifying vulnerability and deepening poverty traps (Hallegatte & Rozenberg, 2017).

VI.2. THE NEED TO UPDATE RISK MANAGEMENT TECHNIQUES IN THE MICROFINANCE INDUSTRY, INCLUDING FINANCIAL CLIMATE RISK

The literature(Beck, n.d.; Brau & Woller, 2004; Hermes & Lensink, 2007) on financial stability focuses on the study of financial entities deemed "too big to fail" due to their significant impact on the performance of economies. There is a notable parallel in the microfinance literature. Microfinance institutions (MFIs) can be likened to these too-big-to-fail entities, given their crucial role in providing financing to various levels of the lower-income population.

Risk management in MFIs necessitates an update in their risk management techniques to properly and explicitly identify and manage financial climate risks. This is especially relevant as the convergence of climate change, finance, and microfinance creates a vital domain for enhancing risk management methods in MFIs. Climate change introduces additional threats and unpredictability, particularly for individuals in poverty and small enterprises, who are the primary clientele of microfinance(Fenton et al., 2015). The risks in this context include but are not limited to, increased frequency and severity of natural disasters, challenges in agricultural production, economic volatility, as well as cascading or compounding events as discussed earlier. Thus, it is imperative for MFIs to promptly adapt their risk management systems to efficiently mitigate these escalating risks.

In this context, it is critical to prioritize the implementation of climate risk assessments. This involves integrating climate models and projections into the risk management frameworks of MFIs to anticipate and prepare for potential impacts on borrowers' loan repayment capacities.

Additionally, the development of financial products specifically designed to address the needs of those affected by climate change, such as insurance products that protect farmers from crop failure due to extreme weather conditions, could offer a protective measure and lessen the economic vulnerability of low-income groups.

Lastly, leveraging technology to improve risk management is crucial. For example, utilizing mobile technology for immediate data collection and analysis could enhance MFIs' ability to monitor climate change-related risks more efficiently and enable them to make informed financing decisions.

VII.    CONCLUSION

The document delves into the challenges and advancements in defining financial climate risk in the context of the impacts of climate change on the economy and financial systems.

The analysis expored the concept of risk by the IPCC, , and economic climate impact triggered by the IPCC and developed in the Stern Review, as well as some of the new literature in financial climate risk. A finding of this assessement is in the communication of climate change risk and financial climate risk targeted to larger readerships. The mass communication of risk is less accurate in the use of concept of risk, than that of  scientific research, which provide a more transparent description of the meaning of risk for the authors.

It further discusses the evolution of risk conceptualization and the hurdles in defining financial climate risk, highlighting the necessity to consider the compounding and interconnected nature of climate risks.

Moreover, the document sheds light on the shift towards methodologies and models that account for risk composition, alongside efforts to weave climate change into economic and financial theories.

It also underscores the role of entities like the Network for Greening the Financial System (NGFS) in evaluating climate impacts at macroeconomic and financial system levels, as well as the demand for comprehensive tools to weave compound risks into climate scenario analysis.

At the same time, it points out the need for more nuanced scientific research, as current studies predominantly focus on whole economies, economic sectors, financial markets, and public financial market participants. Yet, considerable untapped research potential exists within medium to small enterprises at the nascent stages of development.

The document stresses the importance of enhancing risk management strategies in microfinance to safeguard the financial stability of low-income individuals and small businesses against climate change. This requires a holistic approach that includes assessments of climate risks, tailored financial instruments, and technology adoption. Through these measures, Microfinance Institutions (MFIs) can significantly bolster the resilience of vulnerable groups to the adverse effects of climate change.

In summary, the document offers a detailed exploration of the complexities, advancements, and shifting paradigms in defining and tackling financial climate risk within the climate change framework and its repercussions on the economy and financial systems.